# Stability Analysis of Continuous-Time Linear Time-Invariant Systems

Kam Modjtahedzadeh
Department of Physics, UC Santa Barbara

**Abstract**

This paper focuses on the mathematical approaches to the analysis of stability that is a crucial step in the design of dynamical systems. Three methods are presented, namely, absolutely integrable impulse response, Fourier integral, and Laplace transform. The superiority of Laplace transform over the other methods becomes clear for several reasons that include the following: 1) It allows for the analysis of the stable, as well as, the unstable systems. 2) It not only determines absolute stability (a yes/no answer), but also shines light on the relative stability (how stable/unstable the system is), allowing for a design with a good degree of stability. 3) Its algebraic and convolution properties significantly simplify the mathematical manipulations involved in the analysis, especially when tackling a complex system composed of several simpler ones. A brief relevant introduction to the subject of systems is presented for the unfamiliar reader. Additionally, appropriate physical concepts and examples are presented for better clarity.

## Introduction

*Stability* is a vast subject in the field of signals and systems, in general, and, controls, in particular. All controlled dynamical systems, being biological, chemical, physical, nuclear, or of any other domain, must have a good degree of stability. The Three Mile Island and Chernobyl catastrophes that occurred due to the process's temperatures going out of control are still fresh in our memories.

Seldom, dynamical systems are intentionally made unstable. That adds an extra layer of difficulty in the design of the systems for acceptable overall stability [5]. Examples are Segway, essentially an inverted pendulum for fun and space efficiency, and jet fighter planes, for high maneuverability.

Considering the degree of complexity of today's systems, stability issue is a huge liability and a headache for the control engineer; hence, the need for relevant and powerful mathematical tools.

A lot has been developed by control theoreticians [4] such as earlier works of Bode, Nichol, and Nyquist, building upon Laplace transform, and later work of Kalman. Current research includes $H^\infty$ (H infinity) and Hardy space [6]. However, coverage of such advanced methods is beyond the scope of this paper.

## System Classifications

Strictly speaking, there are no *linear-time invariant (LTI) systems* in reality. Elements of the systems are functions of the dependent, independent (e.g., time), or both variables. However, such effects become minimal when the systems operate near their equilibrium points, and temperature variation and other environmental effects are kept within a range. Thus the *LTI models* of the systems results in ordinary differential equations (ODE) with constant coefficients.

This paper only concerns the *continuous-time LTI systems*. Although digital controllers have been replacing the analog ones, the concepts developed for continuous domain are extendable to the discrete domain in the framework of z-transform [3], [4].

Another classification of the systems is *causality (causal versus non-causal)*. Almost all real systems are causal as the output depends on the past and present inputs but not on the future inputs. However, time advance (positive time shift), such as preview, and anticipation is actually employed in practice and is non-causal.

## Impulse Response and Transfer Function

As the name implies, the *impulse response*, $h(t)$, of an LTI system is defined as the output when the input is a unit-impulse function, $u(t)$. Practical impulse resembling inputs include, force and voltage gendered by a hammer stroke and fast on-off switching respectively.

The *transfer function*, $H(s)$, of an LTI system is defined as the Laplace transform of the impulse response, with all the initial conditions (IC) set to zero. Alternatively, it can be defined as the Laplace transform of the output over the Laplace transform of the input, when $IC = 0$.

Laplace transform is covered in a dedicated section later in this paper. However, some of its basic operations are employed in this section, in order to accomplish rudimentary system concepts all at once in the beginning.

Consider the RC circuit of figure 1 when a voltage source, $v(t)$, is placed in the loop, as the input, and the capacitor charge, $q(t)$, as the output. The parameters are constants for the range of operation and with respect to time. The voltage source provides a continuous signal. The system is causal as it is a real practical circuit.

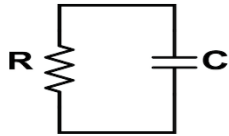

*Fig. 1, RC Circuit*

The corresponding ODE and Laplace transform equations, when $IC = 0$, are:

$$R\frac{d}{dt}q(t) + \frac{1}{C}q(t) = v(t)$$

$$RSQ(s) + \frac{1}{C}Q(s) = V(s)$$

The impulse response and the transfer function are obtained according to their definitions:

$$h(t) = \frac{1}{R}e^{-at}u(t), \quad a = \frac{1}{RC}$$

$$H(s) = \mathfrak{L}\{h(t)\} = \frac{1}{R}\left(\frac{1}{s+a}\right), \quad IC = 0$$

Also, $H(s) = \frac{Q(s)}{V(s)} = \frac{1}{R}\left(\frac{1}{s+a}\right), \quad IC = 0$

An alternative name for the transfer function is *system function*. Note that the impulse response and the transfer function characterize the system and are neither input nor IC dependent; hence the name system function.

A transfer function is a *rational function* when both the numerator and the denominator are polynomials. The roots of the denominator and the numerator are called *poles* and *zeros* respectively. The denominator polynomial set to zero forms the *characteristic equation* of the system (the roots, or poles, determine the types of functions in the partial fraction expansion).

Summarizing the system of figure 1, it is a causal continuous-time LTI system with a rational transfer function. Throughout the rest of this paper, the assumptions of continuous-time linear time-invariant will remain.

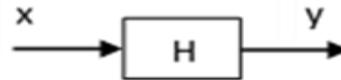

*Fig. 2, Block Diagram of a Basic System*

Systems are often depicted by their corresponding block diagrams, showing their transfer functions, impulse responses, or even Fourier integrals in the blocks. For instance, figure 2 represents a simple system such as that of figure 1.

The solution for the output, $y(t)$, of figure 2 via impulse response is through the convolution integral of the input, $x(t)$, and the impulse response, $h(t)$:

$$y(t) = \int_{-\infty}^{\infty} x(t-\tau)h(\tau)d\tau$$

$$= \int_{-\infty}^{\infty} x(\tau)h(t-\tau)d\tau$$

The integration becomes cumbersome for complicated functions. A much more efficient and systematic solution is by the Laplace transform method, applying its convolution property:

$$Y(s) = H(s)X(s)$$

Inverse Laplace transform can be performed subsequently with ease, with the aid of partial fraction expansion and an inverse table. However, the concept behind convolution is insightful, and also is employed in derivations and proofs.

## Absolutely Integrable Impulse Response

*Bounded input bounded output (BIBO) stability definition* has been vastly accepted as the criterion for the stability of systems [3], [4]. It states that: a system is stable, if for every bounded input the corresponding output is bounded. In other words, the system is stable if the output is finite for all possible finite inputs.

For the particular case of continuous-time LTI systems, it can be proven that *a system is (BIBO) stable, if and only if, the impulse response $h(t)$ is absolutely integrable*.

$$\int_{-\infty}^{\infty} |h(t)| dt < \infty$$

The sufficiency part of the proof is obtained by manipulating the absolute value of the convolution integral:

$$|y(t)| = \left| \int_{-\infty}^{\infty} h(\tau)x(t-\tau)d\tau \right|$$

$$\leq \int_{-\infty}^{\infty} |h(\tau)x(t-\tau)d\tau|$$

$$\leq \int_{-\infty}^{\infty} |h(\tau)||x(t-\tau)|d\tau$$

$$\leq B_x \int_{-\infty}^{\infty} |h(\tau)|d\tau \leq B_y$$

Where $B_x$ and $B_y$ are the input and output bounds respectively and are positive numbers.

For the necessity part of the proof, it can be shown that if the integral is not bounded, then there exists at least a bounded input that will drive the output out of bound.

$$x(t) = sgn(h(-t)) = \begin{cases} 1 & h(-t) > 0 \\ 0 & h(-t) = 0 \\ -1 & h(-t) < 0 \end{cases}$$

$$y(t) = \int_{-\infty}^{\infty} h(\tau)x(t-\tau)d\tau$$

$$y(0) = \int_{-\infty}^{\infty} h(\tau)sgn(h(\tau))d\tau$$

$$= \int_{-\infty}^{\infty} |h(\tau)|d\tau = \infty$$

As an example, the exponentially decaying impulse response of the RC circuit is absolutely integrable and stable. However, had the exponent of the impulse response been positive, then the system would be neither absolutely integrable nor stable. A negative resistor, i.e., physically a voltage source in phase with the input voltage source, would put energy into the system rather than damping out energy, so causing the exponentially growing output.

Note that the absolutely integrable method only reveals the absolute stability, but not the relative stability.

## Fourier integral

*Spectrum, $X(j\omega)$,* of a continuous-time *aperiodic signal, $x(t)$,* is obtained by the *Fourier integral,* also known as the *Fourier transform of $x(t)$*, if the integral converges:

$$X(j\omega) = \mathcal{F}\{x(t)\} = \int_{-\infty}^{\infty} x(t)e^{-j\omega t}dt$$

The transformation is valid, provided that $x(t)$ can be reconstructed accurately by an integral called the *synthesis* equation [1]:

$$x(t) = \frac{1}{2\pi} \int_{-\infty}^{\infty} X(j\omega)e^{j\omega t}d\omega$$

The synthesis and the spectrum equations form the *Fourier transform pair*. A sufficient (but not necessary) condition for the validity of the transformation is, *if $x(t)$ is square integrable* [1]:

$$\int_{-\infty}^{\infty} |x(t)|^2 dt < \infty$$

Alternatively, $x(t)$ *should be absolutely integrable* [1] (as one of the 3 Dirichlet conditions, where the other 2 are normally satisfied for practical signals and systems).

$$\int_{-\infty}^{\infty} |x(t)| dt < \infty$$

The alternative condition also implies that any stable system possesses Fourier transform of its impulse response.

The RC circuit of Figure 1 has an absolutely integrable, as well as a square integrable, impulse response. Thus, its Fourier transform exists and is:

$$H(j\omega) = \frac{1}{R}\left(\frac{1}{j\omega+a}\right), \ a = \frac{1}{RC} > 0$$

The Fourier transform would not exist for if the exponent of $h(t)$ was positive, since the impulse response would explode for $t \to \infty$, making $h(t)$ neither square integrable nor absolutely integrable. Notice that this system is also unstable.

For the system of Figure 2, with a continuous-time LTI system, it can be proved that the Fourier transform of the output is equal to the product of the transforms of the impulse response function and the input; namely, *the convolution property of Fourier transform [3]*:

$$Y(j\omega) = H(j\omega)X(j\omega)$$

This multiplication property holds for any number of cascaded systems that may exist, which greatly simplifies the math.

$H(j\omega)$ Characterizes LTI systems in frequency domain, as $h(t)$ does the same in time domain. Thus, $H(j\omega)$ is called the *frequency response* of the system. Having $H(j\omega)$ in partial fraction form and referring to the table of Fourier transform, $h(t)$ can be determined with ease. Generally, $H(j\omega)$ is complex and is represented in two parts, by either real and imaginary or magnitude and phase functions or plots.

*Fourier transform can be employed for the analysis of stable LTI systems, only.* Stable systems have absolutely integrable impulse responses, which in turn, result in the existence of their Fourier transforms, and vice versa. Thus, its application is limited only to the stable systems, while Laplace transformation, introduced next, is suited for the analysis of both the stable and the unstable systems. Nonetheless, knowledge of frequency response gained from Fourier integral can come handy for stability analysis when working in frequency domain (i.e., Bode, Nichol, and Nyquist methods not covered here) [4].

**Bilateral Laplace Transform**

Bilateral Laplace transform of a continuous-time signal, $x(t)$, is defined as:

$$X(s) = \mathfrak{L}\{x(t)\} = \int_{-\infty}^{\infty} x(t)e^{-st}dt$$

It converts $x(t)$ into the complex function $X(s)$ of complex variable $s = \sigma + j\omega$, provided that the integral exists.

Comparing the formulas for Laplace and Fourier transforms, it can be seen that

$$\mathfrak{L}\{x(t)\} = \mathcal{F}\{x(t)e^{-\sigma t}\}$$

Thus, Laplace transform is the extension and generalization of Fourier transform. The multiplication term, $e^{-\sigma t}$, adds complexity, but also offers flexibility for dealing with a broader class of functions, including many unstable systems. Also, Laplace transform shares algebraic properties with the Fourier transform, which reduces the complexity of math manipulations. The combination of these two facts make Laplace transform particularly useful for the analysis and design of feedback control systems.

Couple of examples are presented to illustrate some important facts about the Laplace transform. First, consider the function $h(t) = e^{-at}u(t)$, i.e., the system of figure 1, where $R$ is set equal to one for simplicity:

$$H(s) = \int_{-\infty}^{\infty}\left(e^{-at}u(t)\right)e^{-st}dt$$

$$= \int_{0}^{\infty} e^{-(a+\sigma)t}e^{-j\omega t}dt$$

We recognize this as the Fourier transform of $e^{-(a+\sigma)t}u(t)$, which is:

$$H(j\omega + \sigma) = \frac{1}{j\omega+(a+\sigma)}, \ a + \sigma > 0$$

$$= \frac{1}{s+a}, \ \sigma > -a$$

This shows the close relationship between the two transforms.

Comparing the convergence constraints for the Fourier and the Laplace transforms of $e^{-at}u(t)$,

we notice that the former is on, $a$, whereas the latter is on, $\sigma$. Let us consider the case where the system is unstable (i.e., exponentially growing, $a < 0$). Since $\sigma$ can take any real value, it can satisfy the $\sigma > -a$ constraint for the Laplace transform to exist. On the contrary, the Fourier transform does not exist, since, $a$, is a fixed system constant and is negative in this case, contradicting the $a > 0$ constraint.

This is a significant advantage of Laplace over Fourier transform. It means that Laplace transformation of both stable and unstable systems are possible, whereas, Fourier transformation of only the stable systems is possible, but not the unstable ones.

Second, consider $h(t) = -e^{-at}u(-t)$.

$$H(s) = -\int_{-\infty}^{0} e^{-at} e^{-st} dt = \frac{1}{s+a}$$

$$= -\frac{e^{-(s+a)t}}{s+a}\Big|_{-\infty}^{0} = \frac{1}{s+a} - \frac{e^{(s+a)\infty}}{s+a}$$

The last term is finite, only if $Re(s) + a < 0$

$$= \frac{1}{s+a}, \quad \sigma < -a$$

Interestingly, Laplace transforms for the two examples with different time functions turn out to be exactly the same. This means the Laplace transform is not unique. Favorably, we notice that the constraints on $\sigma$ differs for the two examples. Therefore, the Laplace transform, together with its associated region of convergence uniquely represent a function.

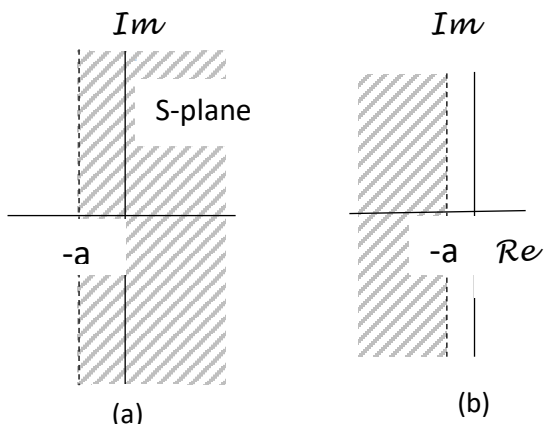

Fig 3, ROC: a) $e^{-at}u(t)$; b) $-e^{-at}u(-t)$

The *region of convergence (ROC)* is the area of the s-plane (i.e., vertical strips) associated with the range of values of, s, that make the Fourier transform integral converge. Figure 3 shows the ROCs for the Laplace transforms of the two examples just presented. They are about a causal (i.e., $h(t) = 0 \ for \ t < 0$) and an anticausal (i.e., $h(t) = 0 \ for \ t > 0$) system, and the associated ROCs are right and left sided respectively. A set of properties with simple proofs, had been developed for the quick determination of ROCs based on the knowledge of $X(s)$ and some characteristics of $x(t)$ [2].

The derivations for the properties are based on the condition for the convergence of Laplace transform [2]. It was pointed out that $\mathfrak{L}\{x(t)\} = \mathcal{F}\{x(t)e^{-\sigma t}\}$, thus, from the condition for the convergence of Fourier transform, the condition for the convergence of Laplace transform is obtained:

$$\int_{-\infty}^{\infty} |x(t)e^{-\sigma t}| dt = \int_{-\infty}^{\infty} |x(t)|e^{-\sigma t} < \infty$$

Three important deductions from the ROC properties relating to this paper [3] are: 1) For a system with a rational system function, causality of the system is equivalent to the ROC being the right half plane to the rightmost pole. 2) An LTI system is stable if and only if the ROC of its system function, $H(s)$, includes the entire $j\omega$-axis. 3) A causal system with rational system function, $H(s)$, is stable, if and only if, all of the poles of $H(s)$ lie in the left-half of the s-plane, i.e., all of the poles have negative real parts.

Convolution property of Laplace transform plays a very important role in the analysis of continuous-time LTI systems, which parallels that of the Fourier transform [3]. For the system of figure 2:

$$Y(s) = H(s)X(s)$$

It maps convolution integral in time domain onto the product of the individual Laplace transform functions in complex domain. For a cascaded system of few, Laplace of the output is simply the product of the Laplace transforms of the input and all of the individual systems in the cascade. This is an extremely powerful property of Laplace transform in simplifying mathematical manipulations. This property, together with the algebraic properties of the transform, make the analysis of complex systems, composed of several simpler ones, easy.

**Feedback Control Systems**

The primary concern in the design of control systems is stability. Consider the block diagram of the simple feedback system, also known as closed-loop system, of figure 4. The first and the second blocks represent the controller and the plant (i.e., system to be controlled) respectively. The output of the system is fed back and compared to a reference (i.e., a scaled desired output). The resulting error feeds the controller and the output of the controller feeds the plant. The controller is the flexible part of the control system that can take any function with any parameters to effectively reduce error over time. The primary objective is to choose the right controller resulting in good stability.

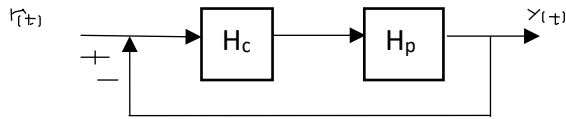

*Fig. 4, A Basic Closed-Loop Control System*

*Stability is a system property* and is independent of the input and ICs. Thus, transfer function alone will reveal the state of system's stability. Employing Laplace transformation, we can write the system of equations for the feedback system, and solve for the overall transfer function by eliminating $E(s)$ in the equations:

$$Y(s) = H_p(s)H_c(s)E(s)$$

$$E(s) = R(s) - Y(s)$$

$$\frac{Y(s)}{R(s)} = \frac{H_c(s)H_p(s)}{1 + H_c(s)H_p(s)}$$

Assume that the plant has an exponentially decaying or growing impulse response, $h_p(t) = e^{-at}$, and the controller is a proportional controller, i.e., $h_c(t) = K\delta(t)$. Thus, $H_p(s) = \frac{1}{s+a}$, $H_c(s) = K$, reducing the transfer function to:

$$\frac{Y(s)}{R(s)} = \frac{K}{s + (K + a)}$$

As it was pointed out in the previous section, a causal system with a rational transfer function is stable, if and only if, all poles of the transfer function (roots of the denominator) lie in the left half of the s-plane. Therefore, the condition for the absolute stability of this system is that the gain, $K$, of the proportional controller has to be larger than, $-a$, regardless of the sign of, $a$:

$$K + a > 0 \;\;\rightarrow\;\; K > -a$$

Therefore, whether the plant of this example is stable or not, we can make the overall feedback system stable by selecting the right range of gain for the proportional controller. The only difference is that the range of $K$ for the unstable as compared to the stable case is smaller (a harder constraint).

Now that the range of $K$ for absolute stability is determined, let us discuss the relative stability, which is also determined from the transfer function. For a causal system with a rational transfer function, the farther the poles to the left of the s-plane, the better stability the system possesses. This is for a simple fact: a system decays faster with larger negative poles, and the faster a system decays, the better relative stability it possesses. By referring to a table of Laplace transform, it can be seen that the real parts of the poles are responsible for the speed of logarithmic decay for the stable systems, and the more negative they are, the faster the system approaches its final value.

By examining the overall impulse response, $H(s) = Ke^{-(K+a)t}$, and system transfer function, $\frac{K}{s+(K+a)}$, for the last example, we can conclude that the larger the $K$: 1) the faster the control system settles, and 2) the farther away to the left, the single real pole is from the $j\omega$ axis, and both of those are signs of better relative stability. The optimum value of K will be determined by combining the stability and performance specification criteria (the latter is not covered here).

We could also employ Fourier integral method for the stability analysis for only when the plant is stable. Parallel to the above development, the overall frequency response would be obtained:

$$H(j\omega) = \frac{K}{j\omega + (K + a)}$$

Then by examining a Fourier integral table we would know that the system is stable only if $K + a > 0$, which is the same exact constraint obtained from the Laplace transform method.

Two short points are mentioned before ending this section. First, control systems are almost always causal. Thus, unilateral Laplace transform could also be applied, where the lower limit of integration is $0^-$ rather than $-\infty$. Second, there is an inverse Laplace integration formula, but it is not used commonly, since solution by partial fraction expansion is simpler.

## Summary

The topic of stability is of paramount importance in the control theory for the obvious reason of safety. The paper began with a brief introduction of systems and related definitions that would be used throughout the paper. It then moved on to the concept of BIBO stability and showed that an LTI system must be absolutely integrable to be BIBO stable. Fourier integral was covered as a tool for the analysis of stable systems. It was pointed out that the frequency response of the systems could be used in frequency domain analysis for stability. Also, significant insight can be gained from Fourier integral in understanding Laplace transform, since Laplace transform is an extension and a generalization of Fourier integral.

Laplace transform was covered as the tool for the analysis of stable as well as unstable systems. From it, both absolute and relative stabilities can be obtained for a system. From pole locations of a causal system with rational system function, absolute as well as relative stability can be determined: A causal system with rational system function is stable if and only if all of the pols have negative real parts. The convolution and algebraic properties of Laplace transform are suited for when dealing with complex systems composed of several connected simpler systems. It is a powerful tool that offers efficient and systematic solutions. In the end Laplace transform method was applied to the analysis of a feedback control system with ease to determine the range of controller parameters for stability.

More advanced techniques are available for the analysis of stability that are beyond the scope of this paper.